\documentclass[aps,jcp,twocolumn,superscriptaddress]{revtex4-1}
\usepackage{graphicx}
\usepackage{amsmath}
\usepackage{amssymb}
\usepackage{pifont}
\usepackage{color}
\graphicspath{{figures/}}

\usepackage{longtable}

\begin{document}

%\title{Double Layer Phenomena Near Charged Semipermeable Surfaces: The role of surface charge and electro-osmotic equilibria}

\title{Electrostatic Interactions and Electro-osmotic Properties of Semipermeable Surfaces. }

\author{Salim R. Maduar}

%\affiliation{A.N.~Frumkin Institute of Physical
%Chemistry and Electrochemistry, Russian Academy of Sciences, 31
%Leninsky Prospect, 119991 Moscow, Russia}

\affiliation{A.N.~Frumkin Institute of Physical
Chemistry and Electrochemistry, Russian Academy of Sciences, 31
Leninsky Prospect, 119071 Moscow, Russia}
\affiliation{Department of Physics, M. V. Lomonosov Moscow State University, 119991 Moscow, Russia}

\author{Olga I.~Vinogradova}

\affiliation{A.N.~Frumkin Institute of Physical
Chemistry and Electrochemistry, Russian Academy of Sciences, 31
Leninsky Prospect, 119071 Moscow, Russia}
\affiliation{Department of Physics, M. V. Lomonosov Moscow State University, 119991 Moscow, Russia}
\affiliation{DWI - Leibniz Institute for Interactive Materials,  RWTH Aachen, Forckenbeckstr. 50, 52056 Aachen, Germany}

\date{\today}
\begin{abstract}
We consider two charged semipermeable membranes, which bound bulk electrolyte solutions and are separated by a thin film of salt-free liquid.  Small counter-ions  permeate into the gap, which leads to a steric charge separation in the system. To quantify the problem, we define an effective surface charge density of imaginary impermeable surface, which mimics an actual semipermeable membrane and greatly simplify analysis. The effective charge depends on separation, generally differ from the real one, and could even be of the opposite sign. From the exact and asymptotic solutions of the nonlinear Poisson-Boltzmann equation, we obtain the distribution of the potential and of counter-ions in the system. We then derive explicit formulae for the disjoining pressure in the gap and electro-osmotic velocity, and show that both are controlled by the effective surface charge.

\end{abstract}
 \maketitle
%\nopacs

%Importance:\\
%It concerns coagulation stability, surface forces and electrokinetic phenomena near semipermeable surfaces and particles which highly depend on surface potential-surface charge relationship. This investigation can provide a basis for interpreting experimental data from SFA, AFM and electrokinetic measurements. If interpreted with assumption of surface impermeability can lead to wrong surface properties.
%
%hypotheses\\
%1)mean-field theory\\
%2)we consider compartments with different concentrations, separated by charged semipermeable membranes in order to account for both surface charge and EO equilibria.
%
%Major findings:\\
%we have shown that semipermeability can dramatically affect double layer phenomena. In certain cases it amplifies surface potential in comparison with impermeable walls leading to stronger electrostatic interaction and fluid flow.
%
%Under other circumstances  (which?) semipermeability results in partial or complete compensation of surface potential which would affect surface force and supress EO flow in some compartments.

\section{Introduction}

Electrostatic Diffuse Layer (EDL) is usually defined as the region where the surface
charge is balanced by the cloud of counterions and local electro-neutrality is not obeyed. It determines both static and dynamic properties of charged objects and results in a variety of phenomena, important for both fundamental and practical applications. Extending over hundreds molecular diameters it results in long-range electrostatic forces between surfaces~\cite{Andelman_biophysics,Nato_andelman,Israelachvili.jn:1992}, which control coagulation stability ~\cite{Electrostat_in_liquids:van_Roij,Hierrezuelo2010,Hang2009,Heijman1995} and open many opportunities  for electrostatic self-assembly~\cite{Bishop2009,Kolny2002,Grzybowski2003,dempster.jm:2016}.  EDL also responds to external electric field, leading to various kinds of electrokinetic phenomena~\cite{Stone2004,Chang2007,Kang2009,maduar.sr:2015,belyaev.av:2011}. The majority of previous work on colloidal forces and electrokinetics has assumed that surfaces are impermeable, so that the EDL profile is determined by the surface charge density and the Debye length of bulk electrolyte solution~\cite{derjaguin.bv:1987,kirby.b:2010}.

The assumption that surfaces are impermeable for ions becomes unrealistic in colloidal systems where membranes are involved. In such cases another factor, surface permeability, comes into play and strongly affects EDLs, so it becomes a very important consideration in interactions involving membranes or determining electrokinetic phenomena. The body of work investigating EDLs near permeable charged surfaces is much less than that for impermeable objects, although there is a growing literature in this area~\cite{nimham.bw:1971,fan.th:2006,jadhao.v:2014}.

\begin{figure}
\includegraphics[width=8cm,clip]{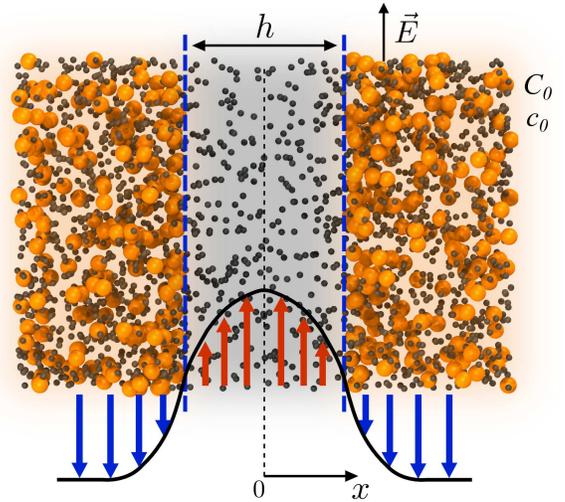}
\caption{Sketch of the bulk solutions of electrolyte bounded by charged semi-permeable membranes and separated by a thin liquid film of thickness $h$.
Membranes are permeable for small counterions only, which partly penetrate to the gap. Large ions remain in the region $\vert x\vert > h/2$. A steric charge separation strongly affects a disjoining pressure, $\Pi (h),$ in the gap. The application of a tangential electric field, $E$, leads to electro-osmotic flows of a solvent (shown by arrows).} \label{fig:capsule}
\end{figure}

Here we  explore what happens when surfaces are semi-permeable, i.e. impermeable for large ions, but allow free diffusion of small counterions. Examples of such surfaces abound in our everyday life. They include bacterial and cell membranes~\cite{sheeler.p:1987}, viral capsids~\cite{Javidpour2013}, liposomes with ion channels~\cite{Meier2000, lindemann.m:2006}, polymersomes\cite{Discher1999,Discher2002:polymerasome_membrane}, free-standing polyelectrolyte multilayer films~\cite{donath.e:1998,vinogradova.oi:2004b,caridade.sg:2013,kim.bs:2006}.  In efforts to understand the connection between EDLs and semipermeability, their formation near membranes has been studied over several years and by several groups~\cite{Maduar2014,vinogradova.oi:2012,maduar.sr:2013,Javidpour2013,Meier2000}. These investigations so far have been limited by the simplest case of electro-neutral membranes and have shown that a steric
charge separation in such a system gives rise to a finite surface potential~\cite{tsekov.r:2008}. This means that due to semi-permeability electro-neutral surfaces demonstrate properties of charged systems~\cite{Maduar2014,vinogradova.oi:2012,lobaskin.v:2012,alan.a:2015,maduar.sr:2013}. In reality, the majority
of semipermeable surfaces are charged. For example, the polyelectrolyte multilayers take a charge of the last deposited polyelectrolyte
layer and the channel proteins determine a charge of biological membranes. However, we are unaware of any previous work that has considered the combined effect of a membrane charge density and a semi-permeability on generation of electrostatic potentials and EDLs.

In this paper we first consider electro-osmotic equilibria between bulk solutions of electrolyte bounded by charged semi-permeable membranes and separated by a thin film of salt-free liquid (see Fig.~\ref{fig:capsule}).  We restrict our consideration to mean-field theory based on the non-linear Poisson-Boltzmann equation (NLPB). We then discuss implications of our theory for the electrostatic interaction of semipermeable membranes and electroosmotic flow in a nanochannel with semipermeable walls.

\section{General Theory}
\subsection{Model}

Consider a solvent confined between two parallel semipermeable membranes at a separation $h$, both are in contact with an electrolyte reservoir. Small (here positive with a charge $z$) ions are free to pass through membranes and leak out from the salt reservoir into the gap, but large (here negative with a charge $Z$, $\vert Z\vert>\vert z\vert$) ions cannot permeate through it. This gives rise a steric charge separation and inhomogeneous equilibrium distribution of ions as sketched in Fig.~\ref{fig:capsule}. Similar system with neutral membranes has been considered before~\cite{vinogradova.oi:2012}.  Now we assume that membranes have a surface charge density, $\tilde \sigma$, which could be either due to dissociation of functional surface groups, or due to the adsorption of ions from solution to the surface~\cite{Andelman_biophysics}.

As before, we use continuum mean-field description by assuming  point-like ions and neglecting ionic correlations. Non-uniform averaged ionic profiles  can then be described by using non-zero electrostatic potential $\phi(x)\equiv ze\psi/k_BT$ and Boltzmann distribution:
$$c_{i,o}(x)=c_{\infty,i,o} \exp(-\phi_{i,o})$$
$$C_{i,o}(x)=C_{\infty,i,o} \exp(-\tilde Z\phi_{i,o})$$
Here $\phi_{i,o}=ze\psi_{i,o}/k_BT$ are the dimensionless electrostatic potentials, and $c_{i,o},C_{i,o}$ are concentrations of small and large ions respectively, where indices $\{i,o\}$ indicate inner ($\vert x\vert <h/2$) and outer ($\vert x \vert > h/2$) solutions.

The NLPB equation for the dimensionless electrostatic potential $\phi$ is then given by
\begin{eqnarray}
& \Delta\phi_o&= - \kappa_i^2 \left( e^{-\phi_o}-e^{-\tilde Z \phi_o} \right) \label{NLPB-1}
\\
& \Delta\phi_i&= - \kappa_i^2\, e^{-\phi_i}
\label{NLPB-2}
\end{eqnarray}
%\begin{eqnarray}
% \Delta\phi= - \kappa_i^2 \left( e^{-\phi}-(H(x-h/2)-H(h/2-x))e^{-\tilde Z \phi} \right) \end{eqnarray}
where the inner inverse screening length, $\kappa_i$, is defined as $\kappa_i^2=4\pi \ell_B c_\infty$ with $\ell_B=z^2e^2/(4\pi\epsilon\epsilon_0k_BT)$ the Bjerrum
length, $\tilde Z=Z/z$ ($<0$) is the valence ratio of large and small ions, and $ c_{\infty}$ is the concentration of small ions at $\vert x\vert \to \infty$. The outer inverse screening length, $\kappa_o$, which represents the inverse Debye length of the bulk electrolyte solution,  can be defined as $\kappa_o^2 = 4\pi \ell_B (\tilde Z^2 C_{\infty} + c_\infty)$, where $C_{\infty}$ is the concentration of large ions at infinity. Since the electroneutrality condition $Z C_{\infty} + z c_{\infty} = 0$ is employed, $\kappa_0 = \kappa_i \sqrt{1 - \tilde Z}$.  We recall that the NLPB has been proven to adequately describe a semipermeable membrane system even at a high valence ratio, $\tilde{Z}=-5$~\cite{Maduar2014}.

To solve Eqs.(\ref{NLPB-1}) and (\ref{NLPB-2}) at the membrane surface, $\vert x\vert = \pm h/2$, we have to impose the boundary condition of the continuity of the potential, and the one of the discontinuity of the electric field:
\begin{equation}
\phi_i'(h/2)-\phi_o'(h/2)=\kappa \sigma,
\label{Eq:BC}
\end{equation}
where $\sigma=\dfrac{4\pi \ell_B \tilde \sigma/e}{\kappa}$ is the dimensionless surface charge density. At the midplane, $x=0$, the electric field vanishes due to symmetry, $\phi'_i=0$. Finally, we set $\phi_o\to 0$ at infinity.

Now it is convenient to define inner and outer diffuse layer charges:
\begin{eqnarray}
\tilde\sigma_{i}^D = \int\limits_0^{h/2}z e c(x)dx,~
\tilde\sigma_{o}^D = \int\limits_{h/2}^{\infty}[z e c(x)-Ze C(x)]dx,
\end{eqnarray}
which satisfy a global electroneutrality condition:
\begin{equation}
\sigma+\sigma_{i}^D+\sigma_{o}^D=0. \label{Eq:electroneutrality}
\end{equation}
It follows from Gauss's theorem that
\begin{equation}
\kappa_i \sigma_{o}^D = \phi_o'(h/2), ~\kappa_i \sigma_{i}^D =- \phi_i'(h/2),\label{Eq:gauss}
\end{equation}
which suggests that Eq.~(\ref{Eq:BC}) is equivalent to Eq.~(\ref{Eq:electroneutrality}). It is therefore always possible to construct imaginary impermeable surfaces with an effective surface charge density $\sigma_{\rm eff}$, which induce the same potential and, therefore, mimic actual semipermeable membranes. Such an effective charge is equal to $-\sigma_{i}^D$ for an inner area, and to $-\sigma_{o}^D$ for an outer reservoir, and fully characterizes electro-osmotic equilibria in the system of real membranes.

\begin{figure*}[tbp]
\centering
\includegraphics[scale=0.6]{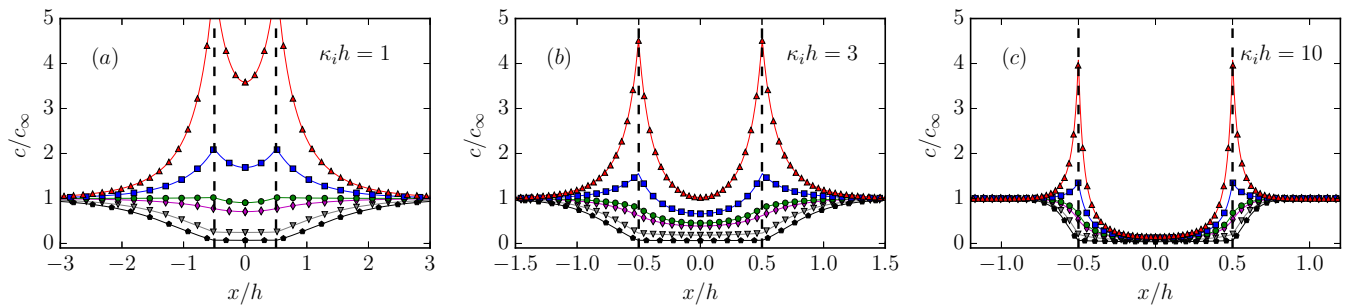}
\caption{Calculated density profiles of ions for (a) overlapping EDLs, $\kappa_i h = 1$, (b) intermediate case, $\kappa_i h = 3$, and (c) non-overlapping EDLs, $\kappa_i h = 10$. From top to bottom $\sigma = -5; -2; -0.5; 0; 2; 5$.  %\textbf{Do we need labels for charge on figures?}
}\label{FIG:concentration}
\end{figure*}

\subsection{Concentration profiles and electrostatic potential}

To illustrate the approach, we begin by studying concentration profiles of small ions obtained by solving numerically Eqs.~(\ref{NLPB-1}) and (\ref{NLPB-2})  for different values of membrane surface charge $\sigma$. Calculation results are shown in Fig.~\ref{FIG:concentration}. We see that away from membranes ($\vert x-h/2\vert \gg \kappa_o^{-1}$) density profiles turn to $c_\infty$. However, in the vicinity of membranes they are generally non-uniform due to EDL formation in both inner and outer regions.    When membranes are close to each other,  $\kappa_i h = 1$ (see Fig.~\ref{FIG:concentration}(a)), inner EDLs strongly overlap. When the surface charge is negative, the counter-ion density, $c_m$, at the mid-plane is finite, and for large negative charges it can be larger than $c_{\infty}$. In other words we observe a counter-ion enrichment in the thin film. In the case of the positive surface charge, $c_m$ is always smaller than  $c_{\infty}$, i.e. we deal with a counter-ion depletion. At a large positive surface charge density the distribution of counter-ions in the gap becomes uniform and even nearly vanishing, which indicates that only outer EDLs are formed to balance the surface charge. When the gap is large, $\kappa_i h = 10$ (see Fig.~\ref{FIG:concentration}(c)), inner EDLs practically do not overlap. We also see that in this case we always observe a counter-ion depletion in the gap. At large positive surface charge counter-ions practically do not diffuse into the gap. Altogether the numerical results presented in Fig.~\ref{FIG:concentration} indicate that formation of EDLs near semi-permeable surfaces
no longer reflects the sole surface charge density.

We remark and stress that the charge of EDLs is not always opposite to the surface charge, as it would be expected for impermeable walls. Since only counter-ions penetrate the gap, so that the inner region can be only positively charged or nearly neutral (if membranes are strongly positively charged as discussed above), $\sigma_{i}^D\geq0$. A vanishing $\sigma_{i}^D$ indicates that inner EDLs disappear and practically all diffuse charges are outside the slit, $\sigma_o^D \simeq -\sigma $. Eq.(\ref{Eq:electroneutrality}) implies that the outer EDL is negatively charged if $\sigma\ge0$. However, for negatively charged membranes the situation can be more complicated than the usual picture. For a relatively small negative surface charge we observe $\sigma_o^D<0$, but for higher negative surface charges $\sigma_o^D$ becomes positive. Therefore, at a certain surface charge, $\sigma=\sigma_0<0$, the outer double layer should fully disappear since all diffuse charges are confined in the slit, $\sigma_i^D = -\sigma $.

\begin{figure}
\centering
\includegraphics[scale=0.6]{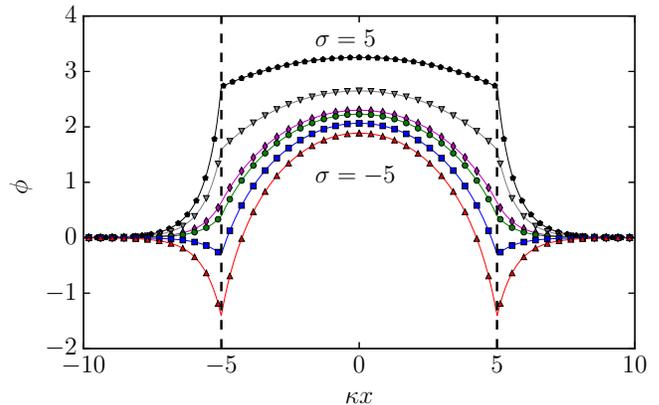}
\caption{A distribution of a potential in the system obtained from NLPB theory at $\kappa h =10$. From top to bottom $\sigma = 5; 2; 0; -0.5; -2; -5$.}\label{FIG:phi}
\end{figure}

The distribution of a potential calculated for a fixed thick film, $\kappa_i h=10$, and different values of $\sigma$ is shown in Fig.~\ref{FIG:phi}. We first remark that  in the case of neutral membranes, $\sigma = 0$, the surface potential, $\phi_s$, is positive, and the distribution of a potential in the system is inhomogeneous. This observation has been reported before~\cite{vinogradova.oi:2012}. The surface potential is always of the same sign as the surface charge for large $\vert \sigma\vert$, but at low values of negative surface charge $\phi_s$ could vanish or become positive.

Let us now use Eq.~(\ref{NLPB-2}) to obtain exact expressions for concentration and potential profiles in the slit. This leads to a Gouy-type expression)~\cite{vinogradova.oi:2012}:
\begin{equation}
\phi_i=\phi_m+\ln\left[ \cos^2\left({\sqrt{2}\over 2} e^{-\phi_m/2} \kappa_i x\right)\right]
\label{phin-ex}
\end{equation}
where $\phi_m$ is the (dimensionless) potential at the mid-plane. By comparing the Gouy solution~\cite{Andelman_biophysics} for impermeable surfaces with  Eq.~(\ref{phin-ex}) we can define the effective surface charge as $\phi_i'(h/2)=\kappa \sigma_{{\rm eff}, i}$ (cf. Eq.~(\ref{Eq:gauss})).
% A similar conception of effective charge (which differs from charge renormalisation) was introduced in \cite{colla.t:2014} to describe equilibrium properties of  %microgels in electrolyte solution.
The dimensionless inner effective surface charge is then:
\begin{eqnarray}
\sigma_{{\rm eff},i}
=-{{\sqrt2}e^{-\phi_m/2}} {\tan\left(\frac{\sqrt2}{2}e^{-\phi_m/2}\kappa_i \dfrac h2\right)},\label{Eq:EffChg}
%\\= -\sqrt 2\sqrt{e^{-\phi_s}-e^{-\phi_m}}\label{Eff_Chg-1}
\end{eqnarray}
and the outer effective charge always differs from the inner and is given by $\sigma_{{\rm eff},o} = \sigma - \sigma_{{\rm eff},i}$. We note that Eq.(\ref{Eq:EffChg}) indicates that effective charges depend on separation between membranes.

%%%%%%%%%%%%%%%%%%%%%%%%%%%%%%%%%%%%%%%%%%%%%%%%%%%%

In many cases properties of the system can be related to $\phi_s$ or $\phi_m$. Therefore, below we focus on their analysis. First we rewrite the differential equations for a potential, $\phi$, into self-consistent algebraic equations for $\phi_s$ and $\phi_m$.
The equation for $\phi_s$ follows immediately from Eq.~(\ref{phin-ex}) by setting $x=h/2$
\begin{equation}
\phi_s=\phi_m+ \ln\left[\cos^2\left( \frac{\sqrt2}{2}e^{-\phi_m/2}\kappa_i \dfrac h2  \right)\right]
\label{SC-2}
\end{equation}

The derivation of the equation for $\phi_m$ requires integration of Eqs.~(\ref{NLPB-1}) and (\ref{NLPB-2}), which gives:
\begin{eqnarray}
{1\over 2\kappa_i^2} \left({\partial \phi_o\over \partial x}\right)^2 &=&  e^{-\phi_o}-1-{1\over\tilde Z} \left( e^{-\tilde Z \phi_o} -1\right)
\label{dphi-o}\\
{1\over 2\kappa_i^2} \left({\partial \phi_i\over \partial x}\right)^2 &=&  e^{-\phi_i}-e^{-\phi_m}
%= \frac{\sigma_{eff}^2}2
\label{dphi-i}
\end{eqnarray}
By setting $x=h/2$ and applying the boundary condition for $\phi'$, Eq.~(\ref{Eq:BC}), and the condition of continuity of $\phi$ we derive the second  relation between $\phi_s$ and $\phi_m$:
\begin{equation}
e^{-\tilde Z \phi_s}= 1-\tilde Z \left(\dfrac{\sigma^2}{2}-\sigma_{{\rm eff}, i}\sigma + 1-e^{-\phi_m}\right)
\label{SC-1}
\end{equation}

\subsection{Asymptotic analysis}

In the general case, the system of Eqs.~\eqref{SC-2} and \eqref{SC-1} should be solved numerically, but in some limits we can derive asymptotic analytical expressions, which relate $\phi_s$ and $\phi_m$ with $\sigma$ and parameters of the system. Below we focus on limits of large and small $\kappa_i h$ and on situations of strong and weak $\sigma$.

\subsubsection{Large $\kappa_i h$}

In a regime, $\kappa_i h\gg 1$, typical for concentrated solutions and/or very thick gap, it is convenient to introduce a new variable $\xi$ by using Eq.~(\ref{SC-2})  as:
\begin{equation}\label{newxi}
\dfrac{\sqrt 2}{2}e^{-\phi_m/2}\kappa_i h/2 = \dfrac{\pi}2(1-\xi)
\end{equation}
Hence the surface potential in Eq.~(\ref{SC-2}) can be expressed as:
\begin{equation}
\phi_s=2\ln \left[ \dfrac{\kappa_i h \sin(\xi\pi/2)}{\sqrt{2}\pi(1-\xi)} \right]
\label{Eq:phis_xi}
\end{equation}
One can easily prove that $\xi$ decays from 1 to 0 with the increase in $\kappa_i h$ from 0 to $\infty$, so that it is small, when $\kappa_i h$ is large. Since $\phi_s$ is bounded by a constant, $\xi$ decays with $\kappa_i h$ as:
\begin{equation}
\xi \simeq \dfrac{e^{\phi_s/2}2  \sqrt{2}}{2  \sqrt{2} e^{\phi_s/2}+\kappa_i h},\label{Eq:xi}
\end{equation}
and the midplane potential reads
\begin{equation}
\phi_m  \simeq 2\ln\left[ {\sqrt{2}\over 2\pi(1-\xi)}\kappa_i h\right] \propto \ln (\kappa_i h)^2
\label{phim-asymptotic}
\end{equation}
Since in this limit $\xi \ll 1$, it can be neglected in the first-order approximation. Then Eq.~(\ref{phim-asymptotic}) reduces to the known result for neutral membranes~\cite{vinogradova.oi:2012}. This suggests that at large $\kappa_i h$ the midplane potential, $\phi_m$, is insensitive to $\sigma$ being controlled mostly by $\kappa_i h$.

When $\phi_m$ is large, we can derive relation between surface charge and surface potential:
\begin{equation}
e^{-\tilde{Z}\phi_s}+\tilde{Z}\sqrt2 e^{-\phi_s/2}\sigma \simeq 1-\tilde{Z}-\tilde{Z}\frac{\sigma^2}{2}, \label{surface_pot:largekh}
\end{equation}
and allows us to construct then the asymptotic solutions for strongly charged surfaces, $\sigma \gg 1$.
For negative surface charges $\phi_s$ is also negative, $e^{-\tilde{Z}\phi_s}\ll1$ and  $e^{-\phi_s/2}\gg1$, which leads to
\begin{equation}
\phi_s \simeq - 2\ln\left[-\dfrac{2(1-1/\tilde{Z})+\sigma^2}{2\sqrt2 \sigma} \right] \propto - \ln  \sigma^2 \label{Eq:LargeH_NegativeCharge}
\end{equation}
For large positive charges and hence positive $\phi_s$ we can use $e^{-\tilde{Z}\phi_s}\gg1$ and $e^{-\phi_s/2}\ll1$ to derive
 \begin{equation}
 \phi_{s}\simeq -\dfrac{1}{\tilde{Z}}\ln\left[1-\tilde{Z}-\tilde{Z}\sigma^2/2 \right] \propto \dfrac{1}{\vert\tilde{Z}\vert}\ln \sigma^2 \label{Eq:LargeH_PositiveCharge}
\end{equation}
In the case of weak charges, $\sigma \ll 1$, one can construct first-order correction to the surface potential of neutral membranes,  $\phi_s^0=-\dfrac{1}{\tilde{Z}}\ln(1-\tilde{Z})$~\cite{vinogradova.oi:2012}, which takes the form
\begin{equation}
\phi_s = \phi^0_s + \sqrt 2 e^{(\tilde{Z}-\frac12)\phi_s^0}\sigma \propto \sigma \label{Eq:LargeH_SmallCharge}
\end{equation}

We remark and stress that in all  cases above $\phi_s$ does not depend on $\kappa_i h$ being a function of only $\sigma$ and $\tilde{Z}$.

These expressions for $\phi_s$ together with Eq.~(\ref{Eq:EffChg}) can be used to calculate the inner effective charge
\begin{equation}\label{effchargeinner1}
    \sigma_{{\rm eff},i} \simeq - {\sqrt 2 e^{-\phi_s/2}},
\end{equation}
and the outer effective charge is then
\begin{equation}\label{effchargeouter1}
    \sigma_{{\rm eff},o} \simeq \sigma + {\sqrt 2 e^{-\phi_s/2}}
\end{equation}
An important point to note that $\sigma_{{\rm eff},i}$ and $\sigma_{{\rm eff},o}$ differ from $\sigma$, but do not depend on $\kappa_i h$. For neutral surfaces inner and outer effective charges have the same absolute value, but are of the opposite sign.

\subsubsection{Small $\kappa_i h$}

Now we investigate the system at $\kappa_i h\ll1$. Such a situation would be realistic for a very dilute solutions and/or very thin gap. The asymptotic analysis can be performed with the procedure described above, although now $1 - \xi$ should be taken as a small parameter. However, in this limit another, a simpler analysis can be used. Since inner diffuse layers strongly overlap, one can easily verify that $\phi_m \simeq \phi_s$ (a difference between these two potentials, $\propto(\kappa_i h)^2/8$, which can be shown by series expansion of  Eq.~(\ref{SC-2})). Eq.~(\ref{SC-1}) then allows us to obtain the relation between the surface charge and potential
\begin{equation}
\sigma^2/2 +e^{-\phi_s}\kappa_i h\sigma /2 =  e^{-\phi_s}-1-\dfrac{1}{\tilde{Z}}\left(e^{-\tilde{Z}\phi_s}-1\right)\label{surface_pot:smallkh}
\end{equation}

For strongly positively charged surfaces $\phi_s$ is positive. This implies $e^{-\phi_s}\ll1$ and Eq.(\ref{surface_pot:smallkh}) reduces then to
Eq.(\ref{Eq:LargeH_PositiveCharge}), so that $\phi_s$ does not depend on $\kappa_i h$.

For strong negative charges $e^{-\tilde Z \phi_s}\ll1$ and we get
\begin{equation}
\phi_s \simeq -\ln\left[\dfrac{2(1-1/\tilde{Z})  + \sigma^2}{2-\kappa_i h \sigma}\right] \propto -\ln \sigma^2 \label{Eq:SmallH_NegativeCharge}
\end{equation}

In the case of weakly charged surfaces the expansion in the vicinity of the solution for neutral membranes~\cite{vinogradova.oi:2012} gives
\begin{equation}
\phi_s \simeq \frac{\kappa_i h + 2\sigma }{2\sqrt{1-\tilde Z}} \propto \kappa_i h + 2\sigma  \label{Eq:SmallH_SmallCharge}
\end{equation}

Finally, by combining the expressions for $\phi_s$ with Eq.~(\ref{Eq:EffChg}) we evaluate inner effective charge, which in this limit depends on $\kappa_i h$
\begin{equation}\label{effchargeinner2}
  \sigma_{{\rm eff},i} \simeq  -\frac{e^{-\phi_s}\kappa_i h}{2}
\end{equation}
%\textbf{(Note that it corresponds with the equation of electrolneutrality: $2\sigma + c_m h=0$)}

Whence an outer effective charge is
\begin{equation}\label{effchargeouter2}
  \sigma_{{\rm eff},o} \simeq \sigma + \frac{e^{-\phi_s}\kappa_i h}{2}
\end{equation}
We emphasize that $\sigma_{{\rm eff},i}$ and $\sigma_{{\rm eff},o}$ are now becoming dependent on $\kappa_i h$. However, since $\kappa_i h \ll 1$, one can conclude that one can roughly consider $ \sigma_{{\rm eff},i} \simeq 0$, and $\sigma_{{\rm eff},o} \simeq \sigma$.

\section{Results and discussion}

\begin{figure}
\centering
\includegraphics[scale=0.6]{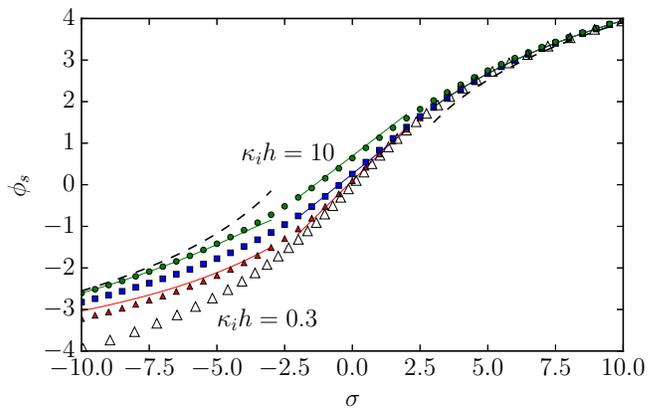}
\caption{Surface potential as a function of surface charge. Filled symbols from top to bottom show numerical results obtained at  $\kappa_i h = 10, 1$ and $0.3$. Solid curves plot predictions of asymptotic solutions given by Eqs.~(\ref{Eq:LargeH_NegativeCharge}) - (\ref{Eq:LargeH_SmallCharge}) and (\ref{Eq:SmallH_NegativeCharge}), (\ref{Eq:SmallH_SmallCharge}).  Dashed curves are calculated with Eq.(\ref{scaling}). Open symbols show results for impermeable charged surfaces,  $\sigma  = 2 \sqrt{2} \sinh(\phi_s/2)$, valid at $\kappa_i h \gg 1$~\citep{Andelman_biophysics}.}
\label{FIG:phis_vs_sigma}
\end{figure}

Here we present the results of numerical solution of  Eqs.(\ref{NLPB-1}) and (\ref{NLPB-2}) and compare them with the above asymptotic expressions.

We begin by discussing $\phi_s$, which has been predicted in general case to
be controlled by $\sigma$ and $\kappa_i h$, and which determines at a given $\kappa_i h$ the effective inner and outer surface charge.
Let us first investigate the effect of $\sigma$ on $\phi_s$ at different values of $\kappa_i h$, which is important for calculating electrostatic interaction energy  as $\int \psi(\tilde\sigma) d\tilde\sigma$~\cite{Behrens1999}. In all calculations we use $\tilde Z = -1$, and vary $\kappa_i h$ from 0.3 to 10. The calculation results are shown in  Fig.~\ref{FIG:phis_vs_sigma}. Also included are numerical results for conventional impermeable walls. We see that $\phi_s$ of membranes significantly differs from the surface potential of impermeable plates of the same $\sigma$, which confirms the important role of semi-permeability. In both cases $\phi_s$ increases with $\sigma$, but the values of $\phi_s$ of membranes are  quantitatively, and even qualitatively different. The only exception is the case of large positive $\sigma$, where numerical calculations show that results obtained at all $\kappa_i h$ converge to a single curve expected for an impermeable wall. We have compared these numerical results with predictions of asymptotic Eq.(\ref{Eq:LargeH_PositiveCharge}), and can conclude that the agreement between numerical results is excellent for all $\kappa_i h$. Remarkably, our results show that Eq.(\ref{Eq:LargeH_PositiveCharge}) is very accurate when $\sigma \geq 2.5$, i.e. its range of applicability is much larger than expected initially. At large negative charge the surface potential increases with $\kappa_i h$. A comparison of asymptotic Eqs.(\ref{Eq:LargeH_NegativeCharge}) and (\ref{Eq:SmallH_NegativeCharge}) with numerical data shows that they are surprisingly accurate when $\sigma \leq -2.5$. Now we recall that all asymptotic expressions for the potential of strongly charged membranes at a given $\tilde Z$ scales as
\begin{equation}\label{scaling}
  \phi_s \propto \pm \ln \sigma^2
\end{equation}
This scaling expression is similar to known for impermeable surfaces~\cite{Andelman_biophysics}. The calculations made with Eq.(\ref{scaling}) are included in Fig.~\ref{FIG:phis_vs_sigma}, and we conclude that they are in agreement with exact numerical results. Eqs.(\ref{Eq:LargeH_SmallCharge}) and (\ref{Eq:SmallH_SmallCharge}), obtained for small charges, are in good agreement with numerical results when  $\vert \sigma \vert \leq 2 $.

%\begin{table*}
%\caption{Surface potential $\phi_s$ for different charges $\sigma$ and distances $\kappa h$. \textbf{Simplify formulae?}}

%\begin{tabular}{c|c|c}
%	\toprule
%    &   $\kappa h \ll 1$ & $\kappa h \gg 1$ \\ \hline
%
% Positive charges
% & $-\dfrac{1}{\tilde{Z}}\ln\left[ 1- \tilde{Z}\left(\sigma^2/2 +1\right)\right]$
% & $-\dfrac{1}{\tilde{Z}}\ln\left[1-\tilde{Z}-\tilde{Z}\sigma^2/2 -\sqrt 2\tilde{Z}\sigma e^{-\phi_{s0}/2} \right]$  \\ \hline
%
%$\vert \sigma \vert < 2$
%& $\dfrac{\kappa h + 2\sigma }{2\sqrt{1-\tilde Z}} $
%& $\phi_s^0+\sqrt 2 e^{(\tilde{Z}-\frac12)\phi_s^0}\sigma$  \\ \hline
%
% Negative charges
% & $-\ln\left[\dfrac{2(1-1/\tilde{Z})  + \sigma^2}{2-\kappa h \sigma}\right]$
% & $-2 \ln\left[\dfrac{2(1-1/\tilde{Z})+\sigma^2}{2\sqrt2\vert\sigma\vert} \right]$  \\ \botrule
%
%\end{tabular}
%\end{table*}

\begin{figure}
\centering
\includegraphics[scale=0.6]{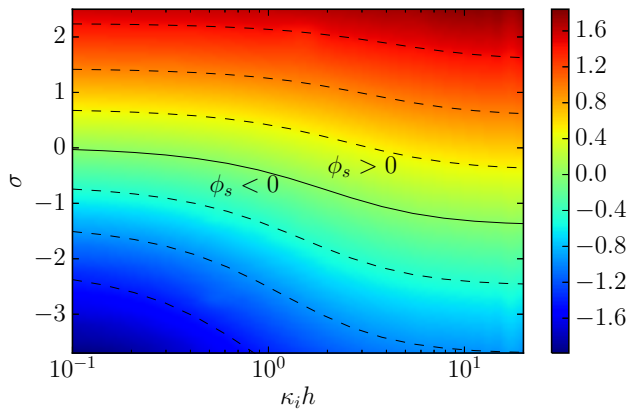}
\caption{Contour lines of $\phi_s$ as a function of $\sigma$ and $\kappa_i h$. The solid curve shows to $\phi_s = 0$, dashed curves show
$\phi_s= \pm 0.5$, $\pm 1$, and $\pm 1.5$.}
\label{FIG:3Dphis}
\end{figure}

Fig.~\ref{FIG:3Dphis} represents a contour plot of $\phi_s$ as a function of $\sigma$ and $\kappa_i h$. We see that for semi-permeable membranes the curve of $\phi_s=0$ generally does not correspond to $\sigma=0$, as it would be expected for impermeable surfaces (except some specific and more complex than considered here cases~\cite{Behrens1999,nimham.bw:1971}). The (negative) charge of zero surface potential decreases from $\sigma \simeq - \kappa_i h$ at small $\kappa_i h$ down to $\sigma \simeq -\sqrt2$ in the limit of large $\kappa_i h$, which can be easily obtained by using Eqs.~(\ref{surface_pot:smallkh}) and (\ref{surface_pot:largekh}). We emphasize that as follows from Eqs.(\ref{effchargeinner1}), (\ref{effchargeouter1}) and (\ref{effchargeinner2}), (\ref{effchargeouter2}) at $\phi_s = 0$ the inner effective charge is equal to $\sigma$, and the outer effective charge vanishes. In other words, only inner diffuse layers are formed. This conclusion is valid for any $\kappa_i h$ as validated by numerical calculations (now shown).

\begin{figure}[ht]
\centering
\includegraphics[scale=0.6]{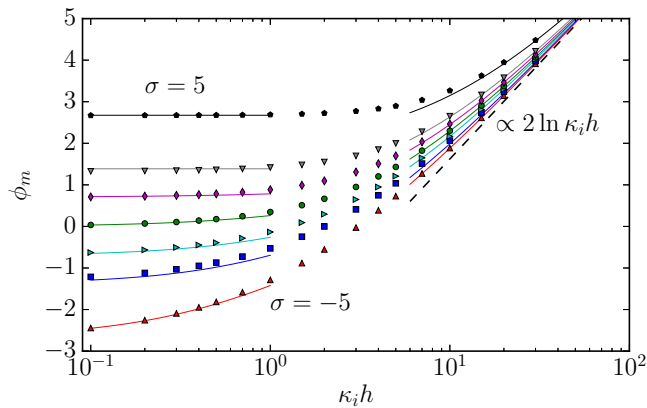}
\caption{The midplane potential as a function of $\kappa_i h$. From top to bottom $\sigma= 5,2,1,0,-1,-2$ and $5$. Symbols show numerical results. Solid curves plot asymtotic results obtained with Eq.~(\ref{phim-asymptotic}) for large $\kappa_i h$ and with Eqs.(\ref{Eq:SmallH_NegativeCharge}),(\ref{Eq:SmallH_SmallCharge}) for small $\kappa_i h$. Dashed line shows  asymptotic behaviour of $\phi_m$ at $\kappa_i h\gg 1$ %\textbf{Dashed line? Its location?}
\label{FIG:phim}}
\end{figure}

We now turn to the midplane potential. The
numerical results for $\phi_m$ as a function of $\kappa_i h$ obtained at several $\sigma$ and predictions of asymptotic theory are given in
Fig.~\ref{FIG:phim} and are again in a good agreement. We note that at a given $\kappa_i h$ the midplane potential, $\phi_m$, monotonously grows with $\sigma$, which in particular imply that the midplane potential for neutral surfaces exceeds that for negatively charged membranes. At large $\kappa_i h$ the midplane potential,  $\phi_m$, diverges as $\ln(\kappa_i h)^2$ as predicted by  Eq.~\eqref{phim-asymptotic}, and we see that indeed the curves are only slightly affected by small parameter $\xi$ (and by $\sigma$) in this equation. The numerical calculations validate asymptotic results at small $\kappa_i h$ and confirm that in this limit $\phi_m$ depends very strongly on $\sigma$.  In this limit $\phi_m \simeq \phi_s$ vanishes for neutral surface, but it is positive for positively charged and negative for negatively charged membranes. The concenration of (positive) small ions in the slit is uniform and equal to $c_{\infty} e^{-\phi_m}$. Therefore, in the limit of $\kappa_i h \ll 1$  for neutral surfaces the concentration of small ions in the gap coincides with $c_{\infty}$. However, if surfaces are positively charged, this concentration becomes smaller than $c_{\infty}$, i.e. the gap between membranes represents a depletion layer of small ions in the system. Note that in this case $ \sigma_{{\rm eff},i} \simeq 0$, and $\sigma_{{\rm eff},o}, \simeq \sigma$ similarly to neutral surfaces. In contrast, in the case of negatively charged surface small ions tend to accumulate in the gap, and their concentration can significantly exceed that in the bulk electrolyte solution.

\section{Implications of results}

In this section we briefly discuss the implications of the above results to interaction of semi-permeable membranes and electro-osmotic flows near them.

\subsection{Interaction of charged semipermeable surfaces}

\begin{figure}
\centering
\includegraphics[scale=0.6]{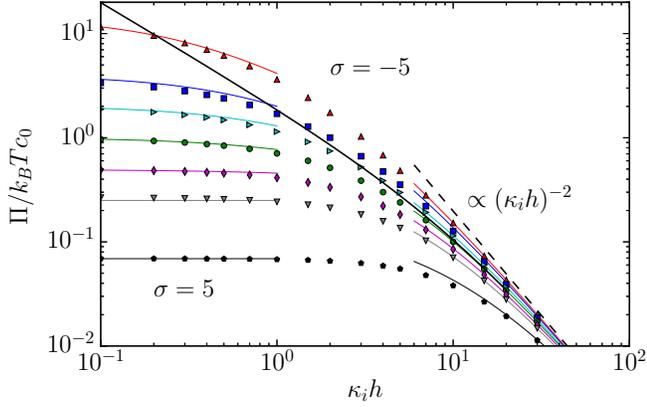}
\caption{Electrostatic disjoining pressure calculated as a function of $\kappa_i h$. Symbols from top to bottom correspond to $\sigma = -5, -2, -1, 0, 1,  2$, and $5$. Solid curve shows the disjoining pressure between impermeable solid walls with $\sigma=-1$. Dashed lines show asymptotic results for large and small $\kappa_i h$ calculated with Eq.(\ref{Eq:disjoining}) by using approximate values for $\phi_m$. }\label{FIG:pressure}
\end{figure}

Since membrane potential depends on $\kappa_i h$, this gives rise to a repulsive electrostatic disjoining pressure in the gap defined as $\Pi \equiv   k_B T (c_{\infty} + C_{\infty})- \Delta p$, where $\Delta p$ is the  force per unit surface on the membrane. We refer the reader to the detailed analysis of $\Delta p$ given in~\cite{vinogradova.oi:2012}, which led to a conclusion that the disjoining pressure can be expressed through the midplane potential as
\begin{equation}
\Pi = k_B T c_{\infty} e^{-\phi_m}\label{Eq:disjoining}
\end{equation}
We can therefore immediately calculate the disjoining pressure as a function of $\kappa_i h$ numerically. The results for different $\sigma$ are shown in Fig.~\ref{FIG:pressure}. As expected, the electrostatic disjoining pressure always decreases with $\kappa_i h$. A startling result is that $\Pi$ decreases with the increase of $\sigma$. This implies, for example, that the electrostatic repulsion of positively charged membranes is always weaker that that of negatively charged and of even neutral membranes. This somewhat counter-intuitive result is a consequence of the behavior of $\phi_m$ discussed above and reflects that small ions accumulate in the gap at a negative surface charge, and strongly deplete when it is positive.

The typical disjoining pressure  curve for the impermeable surfaces calculated using $\sigma = - 1$ is included in Fig.~\ref{FIG:pressure}. It can be seen that the calculations for semi-permeable membranes of the same charge give smaller $\Pi$ at intermediate and especially small $\kappa_i h$. The disjoining pressure does not diverge with a decrease in $\kappa_i h$ ~\cite{Andelman_biophysics,Nato_andelman,Andelman:2010_PB_revisit} by approaching the constant values, which can be easily evaluated using ideal gas approximation and the concept of effective surface charge $\Pi/k_B T c_{\infty}\propto 2\sigma_{\rm eff,i}/\kappa_i h$:
\begin{eqnarray}
 \propto& \sigma^{2 \tilde{Z}}&, \sigma \gg  1\\
\Pi/k_B T c_{\infty} \propto& \, 1-\dfrac{\kappa_i h + \sigma}{\sqrt{1-\tilde Z}} \simeq 1&, \sigma \simeq  0\\
			\propto& \sigma^2   &, \sigma \ll -1
			\end{eqnarray}
Calculations with these equations are shown in Fig.~\ref{FIG:pressure} and are again in excellent agreement with the exact numerical data up to $\kappa_i h \simeq 1$.

At large $\kappa_i h$ the decay of $\Pi$ is not very sensitive to the value of $\sigma$ (see Fig.~\ref{FIG:pressure}). By using asymptotic expression for $\kappa h\gg 1$, Eq.(\ref{phim-asymptotic}), we derive:
\begin{equation}\label{Eq:Pi_large}
\Pi/k_BTc_{\infty} \propto \left(2\sqrt2 e^{\phi_s/2}+\kappa_i h\right)^{-2}
\end{equation}
where the exponential term depends weakly on $\sigma$, which slightly affects results at intermediate $\kappa_i h$. The predictions of Eq.(\ref{Eq:Pi_large}) are included in Fig.~\ref{FIG:pressure}. We can see that this simple analytical result is in good agreement with numerical data. One can conclude that to a leading order $\Pi$ decays as $(\kappa_i h)^{-2}$, which is similar to impermeable surfaces (Gouy-Chapman solution, e.g.~\cite{Xing2011}).

\subsection{Electroosmosis}

\begin{figure}
\centering
\includegraphics[scale=0.6]{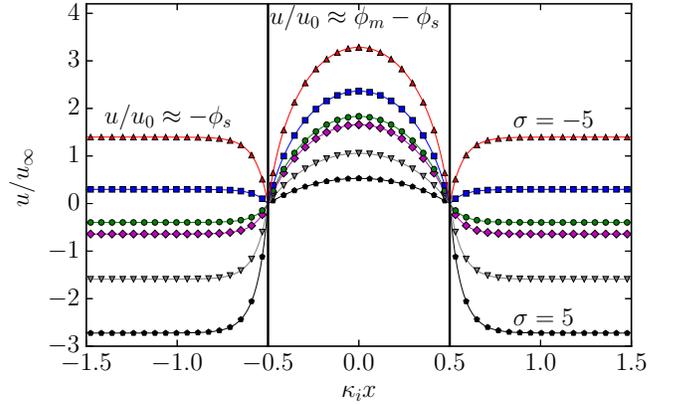}
\caption{Profiles of the electroosmotic velocity obtained  within NLPB theory (symbols) at $\kappa_i h =10$ and different charges $\sigma = -5; -2 ; -0.5; 0; 2; 5$.}\label{FIG:velocity_profile}
\end{figure}

If a tangent electric field, $E$, is applied, the body force in the diffuse layers induces the liquid flow. Now our aim is to relate the velocity, $u$,  of this electro-osmotic flow to $\phi_s$ and $\phi_m$. The liquid flow satisfies Stokes equation:
\begin{equation}
 \eta \Delta  u_{i,o} + \rho_{i,o}  E = 0,\label{Eq:stokes}
\end{equation}
where $\eta$ is dynamic viscosity and $\rho = -\dfrac{\epsilon}{4\pi}\Delta \phi$. By applying the no-slip boundary conditions to  Eq.~(\ref{Eq:stokes}) one can relate $u$ to a potential:
\begin{equation}
u_{i,o}(x)=\frac{\varepsilon E k_BT/e}{4 \pi \eta} (\phi_{i,o} (x)-\phi_s) = u_{0}(\phi_{i,o} (x)-\phi_s)\label{Eq:velocity_solution}
\end{equation}
Here $u_{0} = \frac{\varepsilon E k_BT/e}{4 \pi \eta}$ represents the Smoluchowski electro-osmotic velocity, which would be expected for impermeable surfaces with surface potential $\phi_s=1$.

Eq.(\ref{Eq:velocity_solution}) allows one to calculate electro-osmotic velocity profiles by using the solution for a potential discussed above. The calculation results are shown in Fig.~\ref{FIG:velocity_profile}. A first result emerging from this plot is that in the outer region the electro-osmotic velocity outside of the EDL tends to a constant, which depends on $\sigma$. Its value, $u_{\infty} \simeq -\phi_s u_{0}$, can be easily found from Eq.(\ref{Eq:velocity_solution}). We recall that $\phi_s$ is strongly affected by semi-permeability of membranes, and that it can vanish or even become positive in the case of weakly negatively charged membranes. We see, in particular, that surfaces of $\sigma=-0.5$ (where $\sigma_o^D$ is also negative) induce an outer electro-osmotic flow in the direction opposite to the applied field as it would be for positively charged impermeable surfaces. This example illustrates that the electro-osmotic velocity in the outer region is determined by effective outer charge density of membranes, but not by their intrinsic  charge. Inside the gap the EDL charge, $\sigma_i^D$, is always positive, so that the flow is always in the direction of applied field. Its velocity augments with a decrease in $\sigma$ from 5 (counter-ion depletion) to -5 (counter-ion enrichment). Finally, we note that the mid-plane ($x=0$) velocity can be expressed as $u= u_{0} (\phi_m - \phi_s) $, and that at small $\kappa_i h$ it is negligibly small, but at large gap it increases as $u \simeq 2 u_0 \ln(\kappa_i h)$.

%Inner flow is amplified due to membranes selectivity to ions in solution: only positive ions are allowed to move into the gap where they all move in one direction without opposite dragging force coming from negative ions.

%\iffalse%%%%%%%%%%%%%%%%%%%%%%%%%%%%%%%%%%%%%%%%%%%%%%%%%%%%
%
%{\it Remark:} one could certainly get approximate analytic expression for $\phi_o$ in the limit
%$\tilde Z \gg 1$ since $\exp[-\tilde Z \phi] \gg \exp[-\phi]]$. To be done...
%
%\fi%%%%%%%%%%%%%%%%%%%%%%%%%%%%%%%%%%%%%%%%%%%%%%%%%%%%

\section{Conclusion}

We have examined theoretically electro-osmotic equilibria in a system of two charged semi-permeable membranes separated by a thin film of salt-free liquid. We have shown that these equilibria are fully characterized by an effective surface charge density of membranes we have introduced, which differs from the real surface charge density, and could even be of the opposite sign. Moreover, our model has predicted an alteration of the effective charge density during the approach. By using NLPB theory we have obtained accurate asymptotic formulae for surface and midplane potentials, which have been used to calculate the effective membrane charge and to interpret a distribution of counter-ions in the system. Finally, we have derived explicit formulae for the disjoining pressure in the gap and electro-osmotic velocity in the system, and have demonstrated that they both are determined by the effective surface charge density of membranes.

\section*{Acknowledgements}

This research was partly supported by the Russian Foundation for Basic Research (grant  16-33-00861).

\bibliography{membrane} %your .bib file
\bibliographystyle{rsc} %the RSC's .bst file

\end{document}